\documentclass{emulateapj}

\newcommand{\sqdeg}{deg$^2$}
\newcommand{\cosmos}{$(\Omega_{m},\Omega_{\Lambda})_0$}
\newcommand{\VVmax}{$V_{\rm survey}/V_{\rm max}$}
\newcommand{\Vmax}{$V_{\rm max}$}
\newcommand{\Msun}{\hbox{${\cal M}_{\odot}$}}
\newcommand{\ml}{${\cal M}/L$}

\newcommand{\dd}{{\rm d}}
\newcommand{\gauss}{{\cal G}}

\newcommand{\sigr}{\sigma_{\rm r}}
\newcommand{\sigb}{\sigma_{\rm b}}
\newcommand{\muer}{\mu_{\rm r}}
\newcommand{\mueb}{\mu_{\rm b}}
\newcommand{\phir}{\phi_{\rm r}}
\newcommand{\phib}{\phi_{\rm b}}
\newcommand{\comr}{{\cal C}_{\rm r}}
\newcommand{\comb}{{\cal C}_{\rm b}}
\newcommand{\relr}{{\cal R}_{\rm r}}
\newcommand{\relb}{{\cal R}_{\rm b}}

\begin{document}

\title{Quantifying the bimodal color-magnitude distribution of galaxies}
\shorttitle{THE COLOR-MAGNITUDE DISTRIBUTION OF GALAXIES}
\author{Ivan K.\ Baldry$^1$, Karl Glazebrook$^1$,{} 
  Jon Brinkmann$^2$, \v{Z}eljko Ivezi\'{c}$^3$,\\ Robert H.\ Lupton$^3$, 
  Robert C.\ Nichol$^4$, Alexander S.\ Szalay$^1$}
\shortauthors{BALDRY ET AL.}
\affil{
$^1$Department of Physics \& Astronomy, Johns Hopkins University,
       Baltimore, MD~21218.\\
$^2$Apache Point Observatory, P.O.~Box~59, Sunspot, NM~88349.\\
$^3$Princeton University Observatory, Princeton, NJ~08544.\\
$^4$Department of Physics, Carnegie Mellon University, 5000 Forbes Avenue,
  Pittsburgh, PA~15232.
}
\slugcomment{Accepted 2003 September 25 for publication in ApJ.}

\begin{abstract}
  We analyse the bivariate distribution, in color versus absolute magnitude
  ($u-r$ vs.\ $M_r$), of a low redshift sample of galaxies from the Sloan
  Digital Sky Survey (SDSS; 2400\,\sqdeg, $0.004<z<0.08$, $-23.5<M_r<-15.5$).
  We trace the bimodality of the distribution from luminous to faint galaxies
  by fitting double-Gaussians to the color functions separated in absolute
  magnitude bins. Color-magnitude (CM) relations are obtained for red and blue
  distributions (early- and late-type, predominantly field, galaxies) without
  using any cut in morphology.  Instead, the analysis is based on the
  assumption of normal Gaussian distributions in color. We find that the CM
  relations are well fit by a straight line plus a tanh function.  Both
  relations can be described by a shallow CM trend (slopes of about $-0.04$,
  $-0.05$) plus a steeper transition in the average galaxy properties over
  about two magnitudes.  The midpoints of the transitions ($M_r=-19.8$ and
  $-20.8$ for the red and blue distributions, respectively) occur around
  $2\times10^{10}$\,\Msun\ after converting luminosities to stellar mass.
  Separate luminosity functions are obtained for the two distributions.  The
  red distribution has a more luminous characteristic magnitude and a
  shallower faint-end slope ($M^{*}=-21.5$, $\alpha=-0.8$) compared to the
  blue distribution ($\alpha\approx-1.3$ depending on the parameterization).
  These are approximately converted to galaxy stellar mass functions.  The red
  distribution galaxies have a higher number density per magnitude for masses
  greater than about $3\times10^{10}$\,\Msun.  Using a simple merger model, we
  show that the differences between the two functions are consistent with the
  red distribution being formed from major galaxy mergers.
\end{abstract}


\section{Introduction}
\label{sec:intro}

Optical color-magnitude (CM)\footnote{Note that in this paper, by
  `color-magnitude', we always mean `color versus absolute magnitude'.}
diagrams have been used as scientific diagnostics in astronomy since the
pioneering work of E.\ Hertzsprung and H.~N.\ Russell (c.\ 1910).  While the
CM relations for stars are now well established in terms of stellar evolution
theory, the case for galaxies is less clear.  The optical spectra of galaxies
are dominated by the integrated light from stellar populations, and therefore,
the existence of any CM sequence is related to a correlation of galaxy
luminosity with star-formation history (SFH), stellar initial mass function
(IMF), chemical evolution and/or dust attenuation.  In order to separate CM
relations from color-morphology relations, most of the study of CM relations
has concerned galaxies of a similar morphological type.  The principal
relationship between color and morphology \citep{Holmberg58,RH94} is that more
spheroidal-like galaxies (early types) are generally redder than more
disk-like or irregular galaxies (late types).

A color-magnitude relation for spheroidal-like systems was first established
by \cite{Baum59}. The integrated colors become systematically redder going
from globular star clusters, through dwarf elliptical galaxies, to giant
ellipticals. Later, more precise measurements of luminous E+S0 galaxies in
clusters showed a shallow CM relation with a small intrinsic scatter
\citep{Faber73,VS77}. This relation was associated with a
metallicity-luminosity correlation \citep{Faber73,Larson74}.  However,
\cite*{WTF95} showed that a age-luminosity correlation also fit the
spectroscopic data because of the age-metallicity degeneracy. \cite{KA97}
ruled out the correlation with age being the primary effect because the
predicted evolution of the CM sequence with redshift was more than observed in
this case.  Thus, the CM relation for bright E+S0 galaxies has been
established as a metallicity-luminosity correlation. The intrinsic scatter and
slope of this `E+S0 ridge' can be used to place constraints on the
star-formation and merging histories of these galaxies \citep*{BKT98}.

Color-magnitude relations for early-type spirals were established by
\cite{VG77} and for spirals in general by \cite*{CR64,Visvanathan81,TMA82}.
These CM relations are more complicated than for the E+S0 ridge for a number
of reasons. The intrinsic is scatter is larger \citep{Griersmith80} and the
luminosity correlations can be associated with SFH \citep{PdG98}, dust
attenuation \citep{tully98} and/or metallicity \citep*{ZKH94}. It is probable
that for some morphological types and across some ranges of absolute
magnitude, all three effects are significant.

When all morphological types are considered together, the color distribution
of galaxies can be approximated by the sum of two `normal' Gaussian functions,
a {\em bimodal} function \citep{strateva01}.  The bimodality of the galaxy
population has been known qualitatively for some time. Researchers general
consider E+S0 galaxies to be early types and Sa-Sd spirals and irregulars to
be late types, and \cite{TMA82} noted that ``early and late morphological
types occupy separate branches in the color-magnitude diagram''. With the
advent of large spectroscopic redshift surveys, it is now possible to
precisely analyse this color bimodality as a function of absolute magnitude,
for the field population in particular (whereas previously clusters offered
the best opportunity to study CM relations since all the cluster members are
approximately at the same distance).

A natural explanation for the bimodality is that the two normal distributions
represent different populations of galaxies that are produced by two different
sets of processes.  In other words, formation processes give rise to two
dominant populations that have different average colors and/or color
dispersions.  Evidence that the color bimodality is due to this comes from the
clustering analysis of \cite{budavari03}.  When the galaxy population was
divided into four color bins, the two reddest bins showed a similar clustering
strength to each other, as did the two bluest bins, with a sharp transition in
properties between them.  This can be explained if the dominant effect is the
fraction of galaxies that are part of the red or blue normal distributions,
rather than the average color of the galaxies. Galaxies that are part of the
red distribution are more strongly clustered.

\cite{bell03red} used only colors to define a red sequence from a photometric
redshift survey. The bimodality was observed out to a redshift of unity and
the evolution of the red sequence was quantified. In particular, they noted a
build up of stellar mass on the red sequence by a factor of about two since
$z=1$. This is {\em inconsistent} with a scenario where red early-type
galaxies form early in the Universe and evolve passively to the present day,
and it favors scenarios where the red sequence derives from merger processes.

For our color analysis, we use data from the Sloan Digital Sky Survey (SDSS).
The SDSS is unique for studying the CM distribution of low-redshift galaxies
because the survey has obtained over $10^5$ redshifts for $z<0.1$ galaxies
with associated five-color photometry.  An overview of various bivariate
distributions, including CM relations, is given by \cite{blanton03broadband}.
Here, we focus on one particular color and analyse in more detail the
low-redshift distribution of galaxies ($z<0.08$; $u-r$ vs.\ $M_r$). We also
extract luminosity functions for the red and blue distributions (early- and
late-type galaxies), relate our results to stellar mass and consider a merger
explanation for the bimodality.  The plan of the paper is as follows; in
Section~\ref{sec:data}, we describe the SDSS data and sample selection; in
Section~\ref{sec:distribution}, we show the CM bivariate distribution; in
Section~\ref{sec:methods}, we describe our assumptions, aims and the
parametric analysis of the distribution, and; in Sections~\ref{sec:results}
and~\ref{sec:conclusions}, we present our results and conclusions. A simple
merger model is described in the Appendix.

\section{The SDSS Data and Sample Selection}
\label{sec:data}

The Sloan Digital Sky Survey \citep{york00,stoughton02} is a project, with a
dedicated 2.5-m telescope, designed to image $10^4$\,\sqdeg\ and obtain
spectra of $10^6$ objects. The imaging covers five broadbands, $ugriz$ with
effective wavelengths of 355, 467, 616, 747 and 892\,nm, using a mosaic CCD
camera \citep{gunn98}.  Observations with a 0.5-m photometric telescope
\citep{hogg01} are used to calibrate the 2.5-m telescope images using the
$u'g'r'i'z'$ standard star system \citep{fukugita96,smith02}.  Spectra are
obtained using a 640-fiber fed spectrograph with a wavelength range of 380 to
920\,nm and a resolution of $\lambda/\Delta\lambda\sim1800$ \citep{uomoto99}.
In this paper, we analyse a sample of galaxies selected from the SDSS main
galaxy sample \citep[MGS;][]{strauss02} that selects objects for spectroscopic
followup to a limiting magnitude in the $r$-band.

The imaging data are astrometrically calibrated \citep{pier03} and the images
are reduced using a pipeline {\tt photo} that measures the observing
conditions, and detects and measures objects. In particular, {\tt photo}
produces various types of magnitude measurement including: (i) `Petrosian',
the summed flux in an aperture that depends on the surface-brightness profile
of the object, a modified version of the flux quantity defined by
\cite{Petrosian76}; (ii) `model', a fit to the flux using the best fit of a
de-Vaucouleurs and an exponential profile; (iii) `PSF', a fit using the local
point-spread function.  The magnitudes are extinction-corrected using the dust
maps of \cite*{SFD98}.  Details of the imaging pipelines are given by
\cite{lupton01} and \cite{stoughton02}.

Once a sufficiently large area of sky has been imaged, the data are analysed
using `targeting' software routines that determine the objects to be observed
spectroscopically. The MGS has the following basic criteria:
\begin{eqnarray}
               r_{\rm Petro} & < & r_{\rm limit} \\
            \mu_{r,{\rm 50}} & < & \mu_{r,{\rm 50,limit}} \\
 r_{\rm PSF} - r_{\rm model} & > & s_{\rm limit} \: .
\end{eqnarray}
The first equation sets the magnitude limit of the survey.  The second
equation sets the surface-brightness limit ($\mu_{r,{\rm 50}}$ is the mean
surface brightness within the Petrosian half-light radius).  This is necessary
to avoid targeting too many objects that are instrumental artifacts. The third
equation is used for star-galaxy separation.  The limits have been modified
since the beginning of the survey but over most of the survey, they are given
by
$r_{\rm limit}=17.77$, %-
$\mu_{r,{\rm 50,limit}}=24.5$ mag arcsec$^{-2}$ and %-
$s_{\rm limit}=0.3$. %-
The targets from all the samples (others include luminous red galaxies and
quasars) are then assigned to plates, each with 640 fibers, using a tiling
algorithm \citep{blanton03tile}. Details of the MGS selection are given by
\cite{strauss02}.

Spectra are taken using, typically, three 15-minute exposures in moderate
conditions (the best conditions are used for imaging).  The signal-to-noise
ratio (S/N) is typically 10 per pixel (pixels width $\approx1$--2\AA) for
galaxies in the MGS. The pipeline {\tt spec2d} extracts, and flux and
wavelength calibrates, the spectra. The spectra are then analysed by another
pipeline that classifies and determines the redshift of the object.

\subsection{Subsample Selection from the Main Galaxy Sample}

%% results may change if photo version, sample or k-corrections are changed %%

We use a well-defined subsample of the MGS called `NYU LSS sample12' covering
2400\,\sqdeg.  We set limits on the magnitude as follows: $13.5<r<17.5$ over
30\% and $13.5<r<17.77$ over 70\% of the area.  The 17.5-limit corresponds to
earlier targeting when the imaging and targeting pipelines were significantly
different from the 17.77-limit.\footnote{Magnitude measurements in this paper
  were predominantly derived from {\tt photo} v.~5.2.} This produces a sample
of 207654 objects, of which, 94\% have been observed spectroscopically. The
remaining 6\% are primarily missed due to `fiber collisions', which means that
the tiling pipeline is unable to assign fibers due to another target being
less than 55'' away. This is a limit imposed by the plate and fiber
technology. When two or more MGS targets are within 55'' of each other, a
fiber is assigned at random to one of them. Of the spectroscopically observed
targets, 99.5\% have reliable redshifts determined and, of these, 97.7\% are
galaxies with redshifts between 0.001 and 0.3.

We further restrict our sample to a low redshift range of $0.004<z<0.080$ and
a range in absolute magnitude of $-23.5<M_r<-15.5$, given by
\begin{equation}
M_r = r - k_r - 5\log(D_L/{\rm \,10\,pc})
\label{eqn:abs-mag}
\end{equation}
where: $r$ is the Milky-Way extinction-corrected, Petrosian magnitude;
$D_L$ is the luminosity distance for a cosmology with \cosmos\ =
(0.3,0.7) and $H_0 = (h_{70}) {\rm\,70\,km\,s^{-1}\,Mpc^{-1}}$, and;
$k_r$ is the k-correction using the method of
\cite{blanton03kcorr}.\footnote{The $k$-corrections were derived from
  {\tt kcorrect} v.~1.16.} This produces a sample of 66846 galaxies
with reliable redshift measurements.

Including higher redshift galaxies can leverage better statistics on the
bright galaxies but here we are also interested in the continuity between low
and high luminosity galaxies.  In addition, restricting the sample to
$z<0.08$, reduces evolution effects and uncertainties in $k$-corrections.
\cite{blanton03broadband} reduced these types of uncertainties by
$k$-correcting to the $z=0.1$ bandpasses.  This is optimal for the median
redshift galaxies in the MGS but is sub-optimal for low luminosity galaxies
(only observed near $z=0$). Therefore, we keep to the standard definition of
$k$-corrections (to $z=0$). This also means that no extrapolation is required
to get from the observed-frame bandpasses to the rest-frame $u$ and $r$ bands
principally used in our analysis.

\section{The Bivariate Distribution}
\label{sec:distribution}

For a spectral-type indicator, we use the rest-frame $u-r$ color defined by
\footnote{We use the magnitudes as defined by the SDSS software pipelines.{}
  To convert to AB magnitudes: $(u-r)({\rm AB}) \approx (u-r)({\rm SDSS}) -
  0.05$ \citep{abazajian03}, and to convert to Vega magnitudes: $(u-r)({\rm
    Vega}) \approx (u-r)({\rm SDSS}) - 0.85$.}
\begin{equation}
C_{ur} = (u_{\rm model} - k_u) - (r_{\rm model} - k_r) \: .
\label{eqn:color}
\end{equation}
This is used because, even without $k$-corrections, the $u-r$ color has been
shown to be a nearly optimal separator into two color types
\citep{strateva01}.  The $u$-band filter observes flux from below the 4000\AA\ 
break and thus any $u-X$ color is highly sensitive to SFH ($X=g,r,i$ or~$z$).
We determined that using $u-r$ gave the most robust results for the analysis
presented in this paper (though $u-g$ gave a marginally better division by
type for the more luminous galaxies).

Model magnitudes are used because they give a higher S/N measurement than the
Petrosian magnitudes, particularly because the $u$-band flux is generally weak
and aperture photometry includes significant Poisson and
background-subtraction uncertainties. In fact, if Petrosian colors are used,
using the $u$-band may not be optimum. For example, \cite{blanton03broadband}
found that the bimodality was most evident in the $^{0.1}(g-r)$ color.  Note
that SDSS model magnitudes are determined using the best-fit profile obtained
from the $r$-band image and fitting only the amplitude in the other bands.

\begin{figure} %[f]
\epsscale{1.1}\plotone{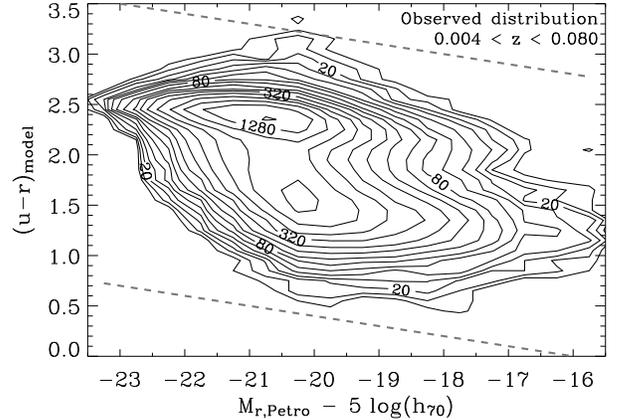}
\caption{Observed bivariate distribution of the sample{} 
  in rest-frame color versus absolute magnitude. The {\em contours} are
  determined for galaxy number counts in 0.1 color $\times$ 0.5 magnitude bins
  (with a total of 66846 galaxies).  The contour levels are on a logarithmic
  scale, starting at 10 and doubling every two contours. The {\em dashed
    lines} represent the limits used in the double-Gaussian fitting described
  in Section~\ref{sec:methods}.}
\label{fig:color-abs-mag1a}
\end{figure}

The bivariate distribution of the sample in $C_{ur}$ versus $M_r$ is shown in
Figure~\ref{fig:color-abs-mag1a}. The bimodality is clearly visible with two
tilted ridges representing the early- and late-type galaxies.  The other $u-X$
CM distributions appear similar (after scaling the color-axis appropriately).
For the $g-X$ CM distributions, the bimodality is still evident (at low
luminosities) but the late-type ridge appears to merge with the early-type
ridge around $M_r\sim-20$ whereas this occurs at slightly higher luminosities
with the $u-X$ colors. This probably reflects the changing dependence of dust
and SFH on the colors of the late-type galaxies. For the remaining CM
distributions ($r-i$, $r-z$, $i-z$), the bimodality is not evident as the
ridges have merged.

\subsection{Correcting for Incompleteness}
\label{sec:completeness}

Before the distribution is analysed, there are two significant incompleteness
issues to deal with:\footnote{We assume that the surface-brightness limit and
  star-galaxy separation criteria do not significantly affect the analysis
  presented here.{} \cite{blanton03broadband} show that the luminosity density
  due to galaxies as a function of surface brightness drops rapidly before the
  limit and \cite{strauss02} determined that only 0.3\% of galaxies brighter
  than an $r$ magnitude of 17.77 are rejected by the star-galaxy separation
  criteria.  In addition, the low redshift sample ($z<0.08$) analysed here
  will be less affected by these selection biases than the majority of
  galaxies in an $r<17.77$ sample (median $z=0.10$).  Instead, the brightest
  galaxies in our sample may suffer from deblending problems. Large galaxies
  are more likely to be blended with foreground stars and they are well
  resolved, which means that {\tt photo} is more likely to measure their
  fluxes incorrectly (by stripping genuine parts from a galaxy).  Their colors
  should be less affected since deblending is applied equally in all bands. We
  also assume that this deblending issue does not significantly affect our
  results. Some discussion of bright-end incompleteness is given by
  \citeauthor{strauss02}} (i) galaxies of a given absolute magnitude and
spectral type can only be observed within a certain redshift range, which in
some cases is much less than the redshift range of the sample, and; (ii) some
galaxies are not observed due to fiber collisions.

To correct for the first issue, we weight each galaxy by a \VVmax\ factor
before recomputing the bivariate distribution, where \Vmax\ is the maximum
volume over which the galaxy could be observed within the sample redshift
range ($0.004<z<0.08$, $V_{\rm survey} = 9.3\times10^6\,{\rm Mpc}^3$).  We
calculate \Vmax\ by iterating to a solution for the $k$-correction at $z_{\rm
  min}$ and $z_{\rm max}$.  The factor, \VVmax, varies from about 1.4 for the
brightest galaxies (set by $r>13.5$), down to 1.01 at $M_r\approx-21$, up to
450/650 for the faintest galaxies (set by $r<17.5$/17.77).  In the 17.77-limit
region, the sample is virtually volume limited between absolute magnitudes of
$-23$ and $-20$ (\VVmax $\la 1.2$). Note also that this correction factor is
principally a function of $M_r$ with little dependence on color at these low
redshifts ($r$-selected sample), which means that this correction is important
for the determination of the luminosity functions but {\em not} for the CM
relations.

The class of galaxies that are not observed due to fiber collisions is not
identical to the whole sample. On average, these galaxies will be found in
higher density regions.  A very similar class of galaxies are those that are
the nearest observed neighbors to the unobserved galaxies. These galaxies
were, predominantly by chance, allocated a fiber instead of their neighbors.
To correct for this issue, we weight these observed galaxies by 2.15.  This
factor is determined from the number of unobserved galaxies divided by the
number of unique nearest observed neighbors, plus unity.

\begin{figure} %[f]
\epsscale{1.1}\plotone{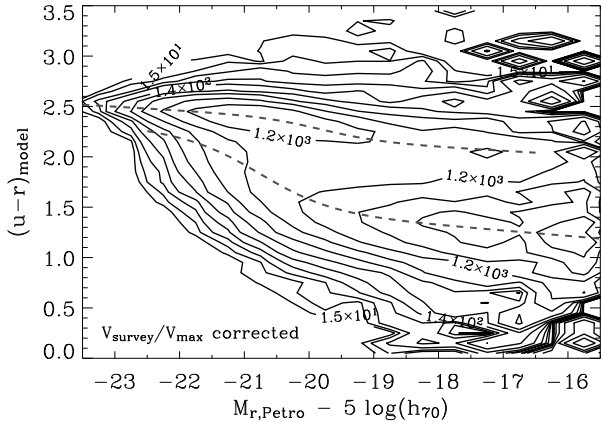}
\caption{Bivariate distribution of the sample{} 
  in rest-frame color versus absolute magnitude, \VVmax\ corrected.  The {\em
    contours} are determined for galaxy number counts in 0.1 color $\times$
  0.5 magnitude bins.  The contour levels are on a logarithmic scale, starting
  at 15 and trebling every two contours.  The upper and lower {\em dashed
    lines} represent a fit to the mean positions of the Gaussian color
  functions for the red and blue distributions, respectively (the fitting is
  described in Sec.~\ref{sec:methods}, see also Fig.~\ref{fig:mean-var}).}
\label{fig:color-abs-mag1b}
\end{figure}

The corrected distribution of galaxies is shown in
Figure~\ref{fig:color-abs-mag1b}. The results of our fitting to the mean color
values along the red and blue distributions are also shown (described later).
In the next section, we describe our parametric fitting to the bimodal
bivariate distribution.

\section{Methodology}
\label{sec:methods}

First of all, we summarize our assumptions and aims before describing our
parameterization and fitting procedure.  Our basic assumptions are given
below.
\begin{enumerate}
\item There are two dominant sets of processes that lead to two distributions
  of galaxies.
\item For each distribution, the average spectral properties vary contiguously
  with visible luminosity. This is reasonable because luminosity is correlated
  with the mass of a galaxy and gravity determines the movement of gas and
  stars.
\item At each luminosity, each distribution can be approximated using a normal
  distribution in the difference between the near-ultraviolet and visible
  magnitudes (a log-normal distribution in the ratio between the fluxes).
  This could result from stochastic variations in SFH, metallicity and dust
  content (and inclination in the case of disks).
\end{enumerate}
Note that for our discussion, we assume that the stellar IMF is universal
\citep{Wyse97,Kroupa02}.

Our aims are:
\begin{enumerate}
\item to quantitatively determine the variation in the mean and dispersion of
  the spectral colors of each distribution, as a function of luminosity;
\item to determine separate luminosity functions;
\item to relate the above to physical explanations;
\item to define a best-fit cut in color versus absolute magnitude space to
  divide galaxies by type.
\end{enumerate}
Our aims differ from other work on early- and late-type galaxies, in that, we
do not use a cut in morphology or spectral type.  Instead, the analysis is
based on the assumption of normal Gaussian distributions.  Nevertheless, we
can safely assume that the red and blue distributions, described in this
paper, correspond in general to the traditional morphological definitions of
early and late types because of the well-known color-morphology relations
\citep[e.g.][]{RH94,shimasaku01,blanton03broadband}.

\subsection{Parameterization}
\label{sec:parameterization}

We assume that the bivariate distribution is the sum of two distinguishable
distributions:
\begin{equation}
\Phi_{\rm comb} = \Phi_{\rm r} + \Phi_{\rm b}
\label{eqn:bimodal-distribution}
\end{equation}
such that $\Phi \,\dd M \, \dd C$ is the number of galaxies between $M_r$ and
$M_r + \dd M$ and between $C_{ur}$ and $C_{ur} + \dd C$.  The parameterization
for these, red and blue, distributions is given by
\begin{equation}
\Phi(M_r, C_{ur}) = \phi(M_r) \, \gauss[C_{ur}, \mu(M_r), \sigma(M_r)]
\label{eqn:distribution}
\end{equation}
where $\phi$ is the luminosity function and $\gauss$ is the color function
parameterized using a Gaussian normal distribution:
\begin{equation}
\gauss(C_{ur}, \mu, \sigma) = \frac{1}{\sigma\sqrt{2\pi}}
  \exp{\left[\frac{-(C_{ur}-\mu)^2}{2\sigma^2}\right]}
\label{eqn:d-gauss}
\end{equation}
Both $\mu$ and $\sigma$ are constrained to be contiguous functions of $M_r$,
in particular, a straight line plus a tanh function given by
\begin{equation}
{\cal T}(M_r) = 
  p_0 + p_1 (M_r + 20) + q_0 \tanh\left[ \frac{M_r - q_1}{q_2} \right]
\label{eqn:sl-tanh}
\end{equation}
This function was found to provide good fits to the data, in particular,
significantly better fits than polynomials with the same number of parameters.
The luminosity functions are fit with \cite{Schechter76} functions that can
be written in terms of magnitudes as
\begin{equation}
\phi(M_r) = 
  c \phi^{*} e^{-c (\alpha + 1)(M_r - M^{*})} e^{-e^{-c(M_r - M^{*})}}
\end{equation}
where: $c \, = \, 0.4 \ln 10$ (=\,0.921034); $M^{*}$ and $\phi^{*}$ are the
characteristic magnitude and number density, and; $\alpha$ is the faint-end
slope.

\subsection{Fitting}
\label{sec:fitting}

For the purposes of fitting to the distribution, the sample was divided into
16 absolute magnitude bins of width 0.5 from $-23.5$ to $-15.5$. Each of these
subsamples was divided into 28 color bins of width 0.1 in $C_{ur}$. The range
in $C_{ur}$ varied from 0.7--3.5 for the most luminous galaxies bin to
0.0--2.8 for the faintest bin, to approximately track the CM relations.

The procedure for fitting to the distribution is given below.
\begin{enumerate}
\item For each absolute magnitude bin, an initial estimate, by eye, was made
  for the mean and dispersion of each distribution.
\item \label{itm:dg-fitting} For each absolute magnitude bin, the distribution
  over the color bins was fitted by a double-Gaussian function with parameters
  $\phir,\muer,\sigr,\phib,\mueb,\sigb$
  (Figs.~\ref{fig:resolve-dg-a}--\ref{fig:resolve-dg-b}).  The fitting used a
  weighted least-squares routine with a grid search in the $\mu$ and $\sigma$
  parameters (narrowing from 0.016 to 0.001).  The variance in each separate
  bin was taken as Poisson (the sum of the galaxy weights squared) plus a
  softening parameter for small number statistics (two times the average
  weight squared at that absolute magnitude)\footnote{The equivalent variance
    for small number statistics with no weighting would be $N+2$ where $N$ is
    the number of measured counts. This provides an approximation to
    uncertainties involved with low counts in a Poisson distribution{}.  This
    expression can be derived by assuming a uniform prior in $N_{\rm true}$
    (the average counts expected, which can be fractional), determining the
    probabilities of measuring $N$ counts for each value of $N_{\rm true}$,
    and finally calculating the probability-weighted mean-square deviation of
    $N_{\rm true}$ from $N$. Using this variance estimate, standard
    least-squares fitting routines can be used with robustness to non-Gaussian
    outliers.} plus a softening for high-count bins (five per cent of the mean
  counts per color bin, squared, at that absolute magnitude).  With these
  additions to the uncertainties, the reduced $\chi^2$ values were on average
  unity.  For the first two absolute magnitude bins ($M_r<-22.5$), $\mueb$ and
  $\sigb$ were not fitted and were fixed at extrapolated values, and for the
  last two bins ($M_r>-16.5$), $\muer$ and $\sigr$ were not fitted.  This is
  because the S/N in these bins was insufficient for a useful 6-parameter fit.
\item ${\cal T}$ functions were fitted to $\sigma$ as a function of $M_r$ for
  each distribution (Fig.~\ref{fig:sigma-var}).
\item Each of the absolute magnitude bins was fitted with double-Gaussian
  functions (as per step~\ref{itm:dg-fitting}) except all the $\sigma$ values
  were fixed by the ${\cal T}$ function fits.
\item \label{itm:t-mu-fitting} ${\cal T}$ functions were fitted to $\mu$ as a
  function of $M_r$ for each distribution (Fig.~\ref{fig:mean-var}).
\item The procedure was repeated up until this point
  (steps~\ref{itm:dg-fitting}--\ref{itm:t-mu-fitting}) until there was no
  significant change in the ${\cal T}$ functions.  This is necessary because
  of the extrapolation described in step~\ref{itm:dg-fitting}. In other
  words, the fitting to the first and last sets of bins depends on the
  extrapolated values.  The result converges quickly in one or two repeats.
\item Each of the absolute magnitude bins was fitted with double-Gaussian
  functions (as per step~\ref{itm:dg-fitting}) except the $\mu$ and $\sigma$
  values were fixed by the ${\cal T}$ function fits.  In other words, only the
  amplitudes of the Gaussian functions were fitted.
\item Schechter functions were fitted to the final luminosity functions
  (Fig.~\ref{fig:lum-funcs}).
\end{enumerate}

To summarize, double-Gaussian functions are fitted to the color functions of
the galaxy distribution divided into absolute magnitude bins.  The dispersions
of the Gaussians are constrained to vary smoothly before refitting the
double-Gaussians, and then the means are constrained.  Alternatively,
constraining the means prior to the dispersions gives a slightly higher total
$\chi^2$ with similar overall results.  The final set of double-Gaussian fits
only allow the amplitudes to vary in order to obtain the luminosity functions
with high S/N.  These are fitted with Schechter functions.

\section{Results and Discussion}
\label{sec:results}

\begin{figure*} %[f]
\epsscale{1.0}\plotone{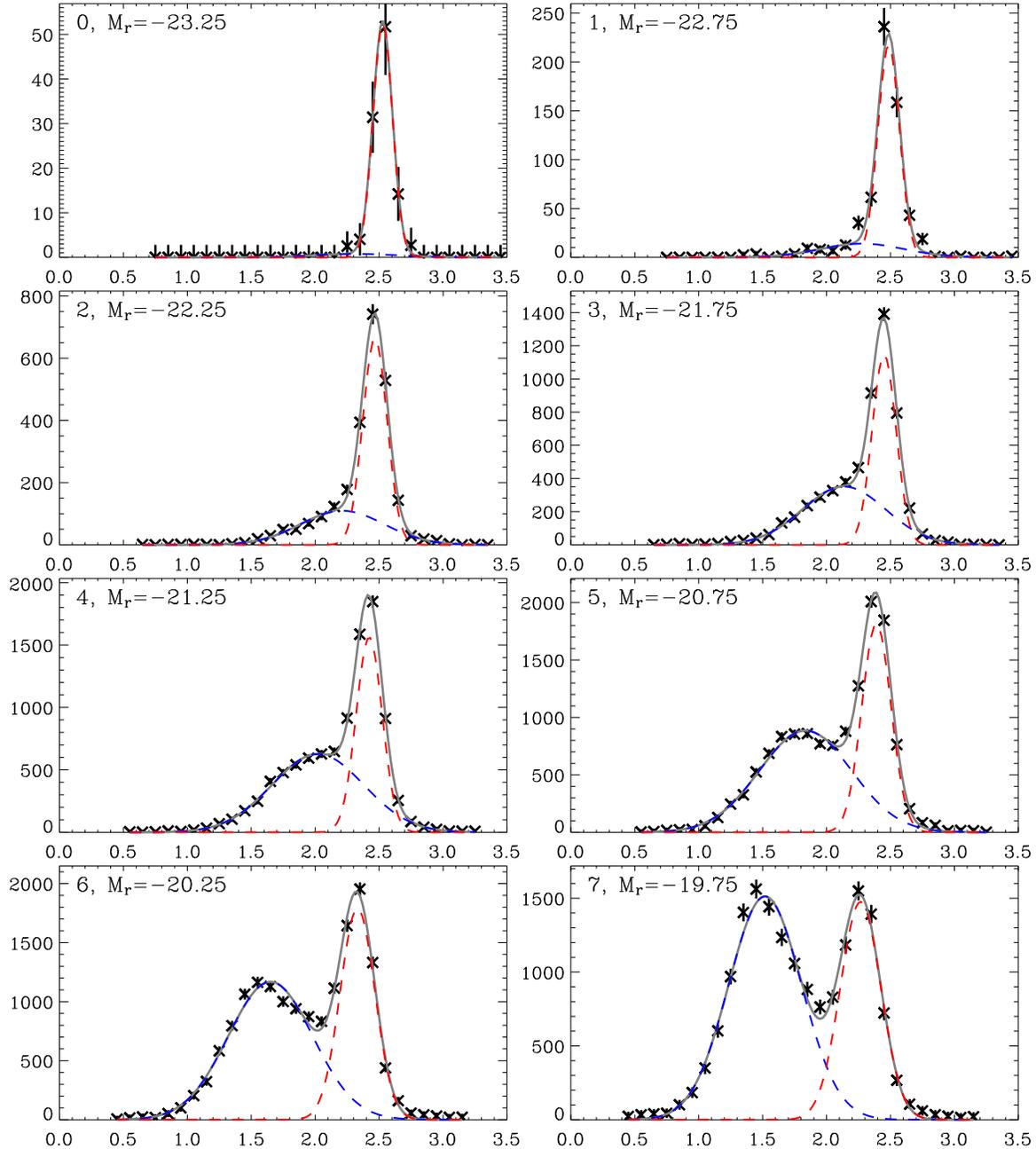}
\caption{Color functions for the galaxy distributions{} 
  in absolute magnitude bins of width 0.5. Each plot shows galaxy number
  counts versus rest-frame $u-r$ color. The {\em crosses} with error bars
  represent the \VVmax\ corrected counts in 0.1 color bins.  The {\em solid
    lines} represent double-Gaussian fits while the {\em dashed lines}
  represent the single Gaussians of the blue and red distributions.}
\label{fig:resolve-dg-a}
\end{figure*}

\begin{figure*} %[f]
\epsscale{1.0}\plotone{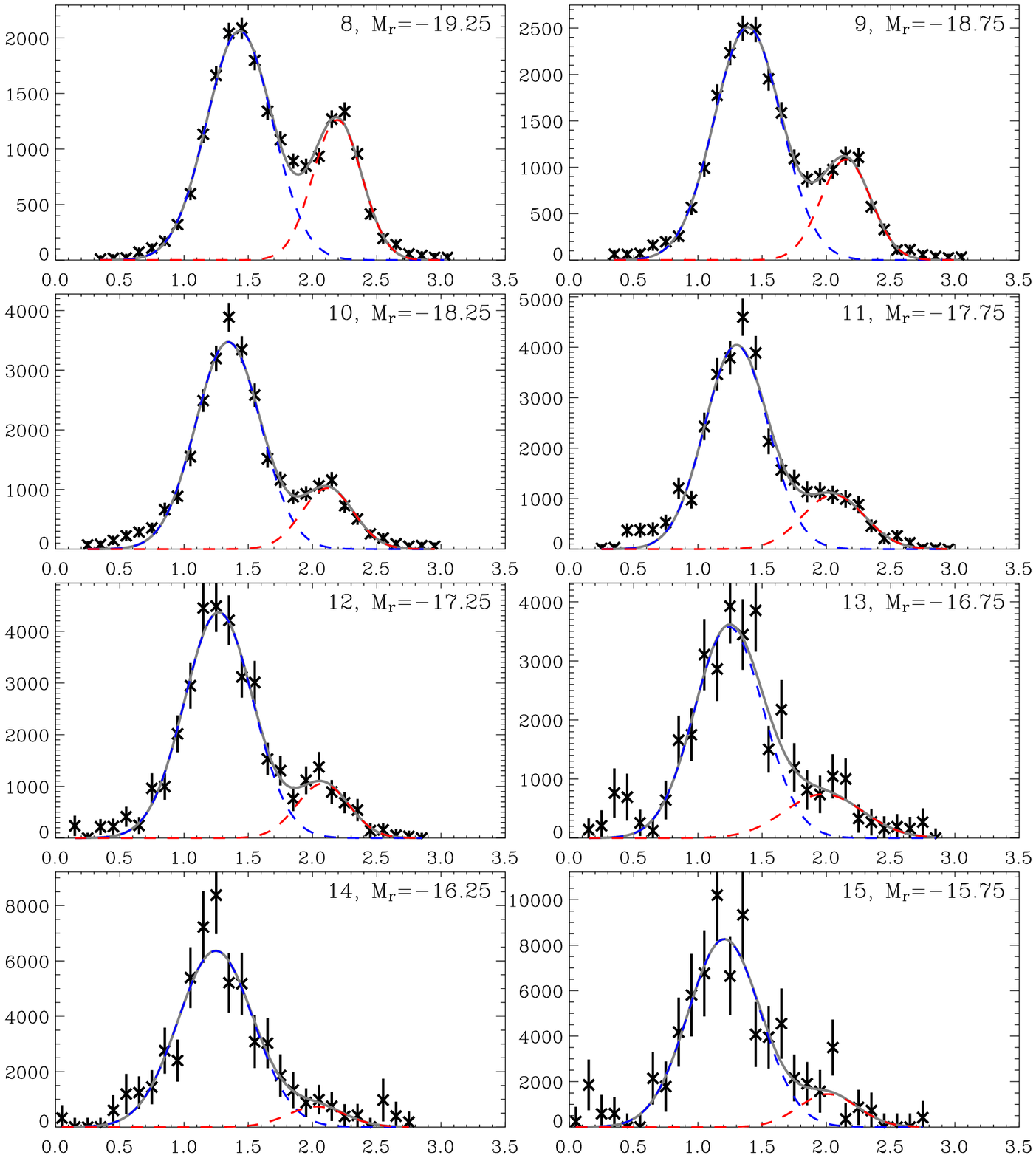}
\caption{Color functions for the galaxy distributions{} 
  {\em continued}.  See Figure~\ref{fig:resolve-dg-a} for details.}
\label{fig:resolve-dg-b}
\end{figure*}

Figures~\ref{fig:resolve-dg-a} and~\ref{fig:resolve-dg-b} show the results of
the double-Gaussian fitting to the color functions.  Visual inspection shows
that the bimodality in the galaxy population is clearly traceable from about
an absolute magnitude of $-22$ to $-17$, and that a double-Gaussian function
provides a good representation for the most part.  For the high S/N mid-range
in $M_r$, there are some significant deviations but with the additional 5\%
systematic uncertainty described in Section~\ref{sec:fitting}, the reduced
$\chi^2$ values are of order unity. We will assume that these slight
non-Gaussian deviations do not affect our results.

For two of the bins brighter than $-22$, there are significantly more galaxies
on the blue side of the red distribution, justifying the continued use of the
bimodal description. For the most luminous bin, there is no evidence of any
blue distribution and we only have an upper limit on the density of
blue-distribution galaxies here.  For the three bins fainter than $-17$, there
are more galaxies on the red side of the blue distribution than the blue side.
Note that for the two brightest and two faintest bins, the mean and dispersion
of the less-populous distribution are fixed by extrapolation from the whole
population. The general trend is for the number-density ratio of the red to
the blue distribution to increase with luminosity.  In the following
subsections, we describe: in \ref{sec:cm-relations}, the CM relations for the
two distributions; in \ref{sec:lum-funcs}, the luminosity functions; in
\ref{sec:divide}, an optimum divider between the two types, and; in
\ref{sec:stellar-mass}, a conversion to stellar mass.

\subsection{Color-Magnitude Relations}
\label{sec:cm-relations}

\begin{figure} %[f]
\epsscale{1.1}\plotone{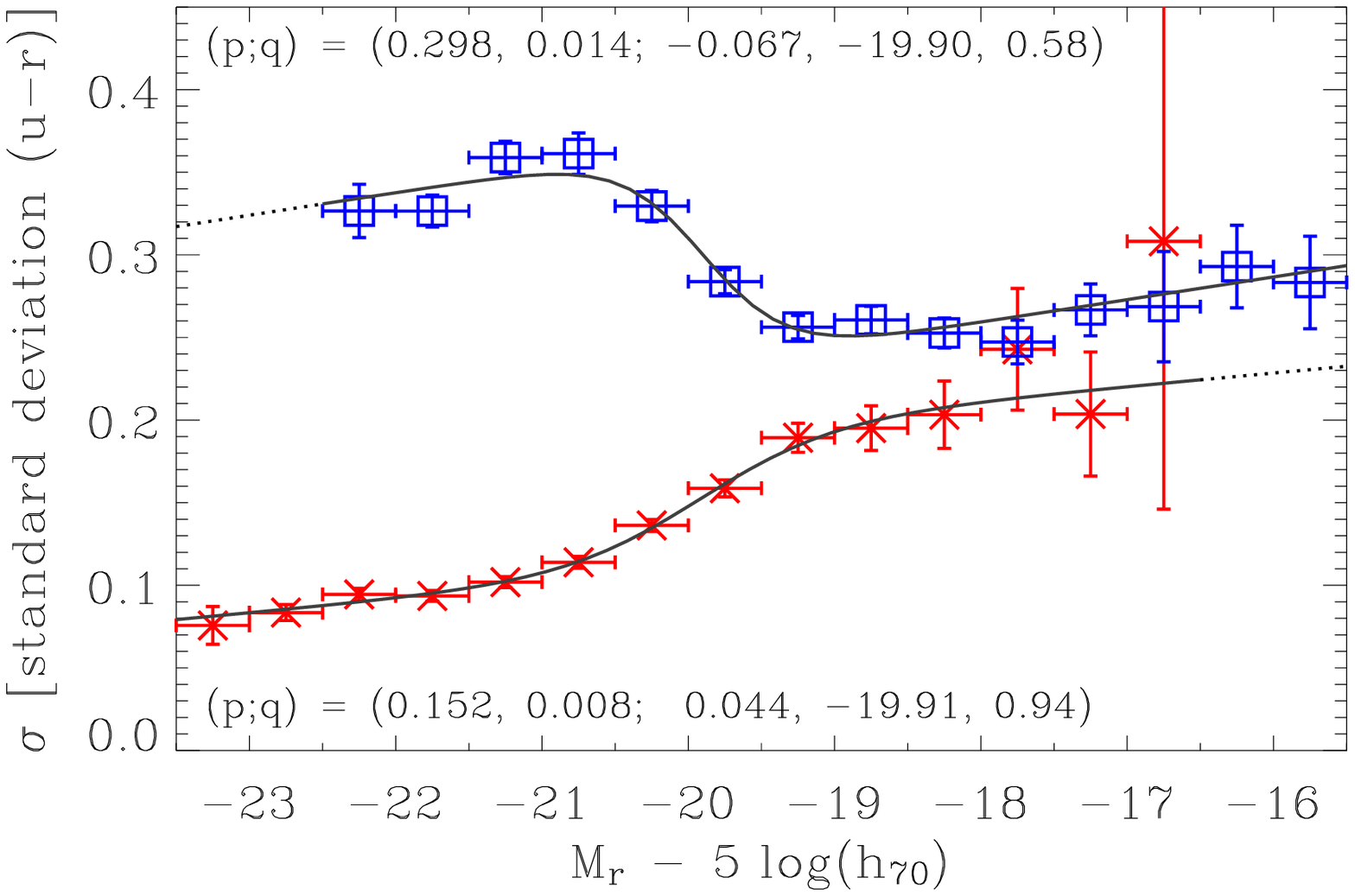}
\caption{Dispersion-magnitude relations:{} 
  variation in the dispersion of the rest-frame $u-r$ colors for each galaxy
  distribution from the double-Gaussian fitting ({\em crosses} for
  $\sigr(M_r)$ and {\em squares} for $\sigb(M_r)$, with vertical error bars
  and horizontal bars representing the width of the magnitude bins).  Note
  that the error bars can be smaller than the symbols.  The {\em solid lines}
  represent straight line plus tanh function (Eqn.~\ref{eqn:sl-tanh}) fits to
  the data.  The {\em dotted lines} represent an extrapolation where the
  parameters are fixed in the double-Gaussian fitting.  The parameters for the
  fitted functions are shown in the plot. Note that the measured dispersion
  includes observational uncertainties.}
\label{fig:sigma-var}
\end{figure}

\begin{figure} %[f]
\epsscale{1.1}\plotone{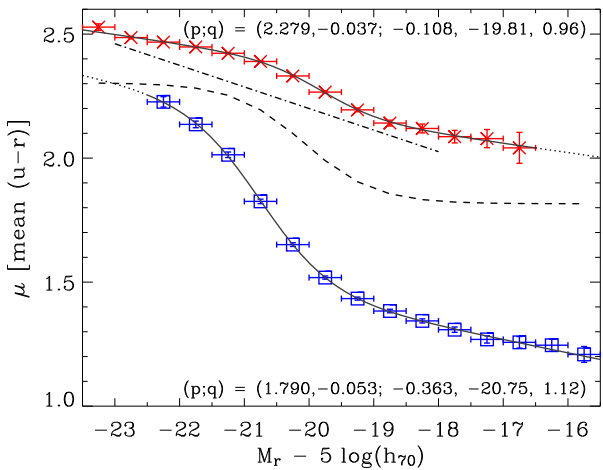}
\caption{Color-magnitude relations:{} 
  variation in the mean of the rest-frame $u-r$ colors for each galaxy
  distribution ({\em crosses} for $\muer(M_r)$ and {\em squares} for
  $\mueb(M_r)$, with vertical error bars).  Note that the error bars can be
  smaller than the symbols.  The {\em dashed line} represents an optimal
  divider between the two distributions (Sec.~\ref{sec:divide}).  The {\em
    dash-and-dotted line} shows the slope of the average $U-V$ CM relation of
  \cite{BLE92B}\ for E+S0 galaxies in clusters ($-0.087$), offset from the red
  distribution for clarity. See Figure~\ref{fig:sigma-var} for other details.}
\label{fig:mean-var}
\end{figure}

To quantify these distributions further, we assume that the Gaussian
parameters vary smoothly from one absolute magnitude bin to the next.  The
dispersion and mean of each distribution are fitted by straight line plus tanh
functions (${\cal T}$ functions, five paras., Eqn.~\ref{eqn:sl-tanh}). These
fits are shown in Figures~\ref{fig:sigma-var} and~\ref{fig:mean-var}. These
${\cal T}$ functions provide a far superior fit than a five-parameter (4th
order) polynomial.\footnote{The difference between{} $\chi^2(\rm polynomial)$
  and $\chi^2({\cal T})$ is (1.4, 7.3, 11.3, 41.5) for
  ($\sigr,\sigb,\muer,\mueb$), respectively.} In addition, they are more
stable for extrapolation (a straight line at the outside limits) and they can
be related more readily to a physical explanation (a general trend with
luminosity plus a transition around a particular luminosity).
Table~\ref{tab:t-params} shows the fitted parameters with uncertainties.  The
parameters $p_0$ and $p_1$ represent the intercept and slope of the straight
line while $q_0$, $q_1$ and $q_2$ represent the amplitude, midpoint and range
of the transition.

%% table values may change if photo version or k-corrections are changed %%

\begin{table*} %[f]
\caption{Color- and dispersion-magnitude relations:{} 
  ${\cal T}$ function parameters\rlap{$^a$}}
\label{tab:t-params}
\begin{center}
\begin{tabular}{crrrrrc}\hline
distribution & $p_0$~~~~~~~ & $p_1$~~~~~~~ 
& $q_0$~~~~~~~ & $q_1$~~~~~~ & $q_2$~~~~~~ & ($q_1/$\Msun)\rlap{$^b$} \\ \hline
$\muer$& $ 2.279\pm 0.006$& $-0.037\pm 0.006$& $-0.108\pm 0.017$&
$-19.81\pm  0.07$& $  0.96\pm  0.16$& 1.8$\times10^{10}$\\
$\sigr$& $ 0.152\pm 0.006$& $ 0.008\pm 0.006$& $ 0.044\pm 0.018$&
$-19.91\pm  0.18$& $  0.94\pm  0.40$& 2.0$\times10^{10}$\\
$\mueb$& $ 1.790\pm 0.014$& $-0.053\pm 0.008$& $-0.363\pm 0.029$&
$-20.75\pm  0.05$& $  1.12\pm  0.10$& 2.6$\times10^{10}$\\
$\sigb$& $ 0.298\pm 0.004$& $ 0.014\pm 0.007$& $-0.067\pm 0.014$&
$-19.90\pm  0.07$& $  0.58\pm  0.19$& 0.9$\times10^{10}$\\
\hline
\end{tabular}
\end{center}
{\small
$^a$~The results of fitting a straight line plus a tanh function
(Eqn.~\ref{eqn:sl-tanh}) to the variations, in the means ($\mu$) and
dispersions ($\sigma$), of the red and blue distributions as a function of
$M_r$ (Eqns.~\ref{eqn:distribution} and~\ref{eqn:d-gauss}).  The $p$
parameters represent the straight line while the $q$ parameters represent the
tanh function.  The fitted lines are shown in Figures~\ref{fig:sigma-var}
and~\ref{fig:mean-var}.  Note that the errors quoted do not include 
systematic uncertainties due to photometric calibration or $k$-corrections.
\vspace{0.1cm}
\newline %-
$^b$~The transition midpoint approximately converted to stellar mass
(see Section~\ref{sec:stellar-mass}).
}
\end{table*}

\subsubsection{The red distribution}
\label{sec:red-distrib}

One of the most well-studied relations is the CM relation for luminous
early-type galaxies
\citep*{VS77,SV78B,SV78C,BLE92A,BLE92B,SS92,TCB01,bernardi03D}.  This
corresponds approximately to the red distribution with $M_r\la-20$
(Fig.~\ref{fig:mean-var}). Our formal slope is about $-0.04$ ($p_1$ for
$\muer$) but we find that the slope gets steeper toward the transition
midpoint at $-19.8$ ($q_1$).

Previous work found slopes of around $-0.1$ for the $u-V$ CM relation
(\citeauthor{VS77,BLE92B,TCB01}).  The difference in slope between $u-r$ and
$u-V$ relations is negligible (\citeauthor{VS77}) and therefore there is a
non-trivial difference between our measured slope and previous work.  The
difference can largely be explained by the use of a $\cal T$-function fit
rather than a straight-line fit because a straight-line fit to our data gives
a slope of about $-0.08$.

Other factors that could contribute to a difference in slope are: field versus
cluster environment; Gaussian color function fitting versus E+S0 morphological
selection and; aperture effects \citep{Scodeggio01,bernardi03D}.  However, the
CM relation has been found to be similar between different environments
(\citeauthor{SV78C,TCB01}) and no significant difference has been found
between E and S0 galaxies in the CM relation (\citeauthor{SV78C}) and,
therefore, all morphological types that are genuinely part of the red
distribution may have a similar relation.  An analysis of the difference
between using SDSS model and other magnitude definitions for the CM relation
is given by \citeauthor{bernardi03D} We note that SDSS model colors are
weighted toward the center of a galaxy and therefore the relations presented
here apply to that weighting \citep[see Sec.~4.4.5.5 of][for model fitting
details]{stoughton02}.

For the color dispersion-magnitude relation (Fig.~\ref{fig:sigma-var}), we
find only a modest slope at the bright end with low statistical significance
($p_1$ for $\sigr$ is about 1 standard deviation from zero).  This is
consistent with the CM relation for $M_r\la-21$ being due to a
metallicity-luminosity correlation \citep{Faber73,KA97} since dust reddening
and SFH correlations could also introduce more scatter.

The dispersion-magnitude relation for the red distribution goes through a
transition at the same magnitude, within the uncertainties, as the CM relation
(Figs.~\ref{fig:sigma-var} and~\ref{fig:mean-var}; Table~\ref{tab:t-params}).
This is consistent with the transition being due to an increasing contribution
from recent star formation with decreasing galaxy luminosity (from $M_r$ of
about $-21$ to $-19$; see also \citeauthor{FS00} \citeyear{FS00}).  The colors
of younger stellar populations are more dependent on their ages than older
populations \citep[see e.g.\ fig.~1 of][]{BKT98}, which implies more
dispersion in a CM relation.  In other words, if there has been on average
more recent star formation in a class of galaxies, then their mean color
becomes bluer and the color dispersion increases for any reasonable variation
in their precise SFHs. However, we cannot rule out the transition also being
caused by a metallicity-luminosity correlation as long as the metallicity
dispersion increases with decreasing galaxy luminosity \citep{poggianti01}.

Note that our measurements of dispersion include observational uncertainties.
At the bright-end, the measured dispersion is about 0.09, which is comparable
to the observational uncertainties and is, therefore, consistent with an
intrinsic dispersion of less than 0.05 \citep{VS77,BLE92B}.

\subsubsection{The blue distribution}
\label{sec:blue-distrib}

Color-magnitude relations for late-type galaxies are also an established
phenomenon \citep{Visvanathan81,TMA82,tully98,Wyse82,PdG98}. Here, we
precisely trace a CM relation over seven magnitudes and find that it is very
well fit by a tanh function plus a straight line (Fig.~\ref{fig:mean-var}). 

For the low-luminosity blue-distribution galaxies ($M_r\ga-19$), we find a
shallow CM relation slope ($-0.05$) that is consistent with a
metallicity-luminosity correlation for the following reasons.  Studies of
late-type galaxies yield a strong metallicity-luminosity relation down to low
luminosities from their emission lines \citep{Garnett02,tremonti03} and from
their stellar content \citep{BdJ00}.  In addition, the general slopes of the
CM relations for the red and blue distributions, defined by the $p_1$ values
(i.e.\ excluding the transition), are approximately the same (within $<2$
standard deviations).  Modest correlations of luminosity with SFH and/or dust
are also possible.

Over the luminosity range from $-19.5$ to $-22$ (increasing galaxy
luminosity), we find a significant reddening of the blue sequence that is too
steep to be explained entirely by a metallicity-luminosity correlation.  This
transition can be explained by a combination of an increase in dust content
\citep{giovanelli95,tully98} and a decrease in recent star formation relative
to the total stellar mass of the galaxy \citep{PdG98}.  These processes will
have opposite effects on the dispersion.  Increased dust content will increase
dispersion, because of the range of reddening associated with different disk
orientations, whereas decreased star formation will decrease dispersion
because old stellar populations vary less in color (c.f.\ the luminous red
distribution). Our interpretation of the dispersion-magnitude relation
(Fig.~\ref{fig:sigma-var}) is then that the dust content increase dominates
the transition from $-19.5$ to $-20.8$ ($\sigb$ increases, $\mueb$ increases)
and that the competing processes approximately cancel from $-20.8$ to $-22$
($\sigb$ decreases slightly, $\mueb$ increases).  This explains why the tanh
fit for $\sigb$ does not coincide with that for $\mueb$
(Table~\ref{tab:t-params}). We take the genuine transition in the properties
of the blue distribution to be that defined by the $\mueb$ fit.

\subsection{Luminosity Functions}
\label{sec:lum-funcs}

\begin{figure} %[f]
\epsscale{1.1}\plotone{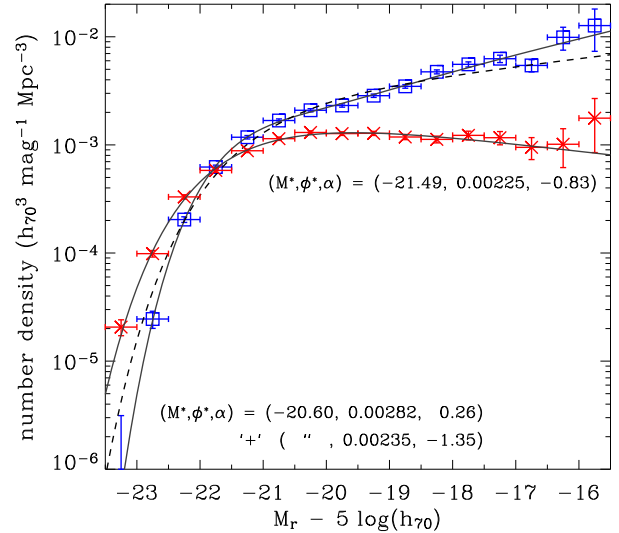}
\caption{Luminosity functions{} 
  for each galaxy distribution ({\em crosses} for $\phir(M_r)$ and {\em
    squares} for $\phib(M_r)$, with error bars).  The lines represent fits to
  the data. The {\em dashed line} for $\phib$ and the {\em solid line} for
  $\phir$ represent standard Schechter functions, while the {\em solid line}
  for $\phib$ represents a double Schechter function with a single value for
  $M^{*}$. The standard single Schechter function does not provide a good fit
  to the blue distribution. The parameters for the single Schechter $\phir$
  and double Schechter $\phib$ fits are shown in the plot.}
\label{fig:lum-funcs}
\end{figure}

The results of fitting the amplitudes of the double-Gaussian functions are
used to determine the luminosity functions (Eqn.~\ref{eqn:distribution}) while
the mean and dispersion of the CM relations are constrained to be ${\cal T}$
functions (Eqn.~\ref{eqn:sl-tanh}). The luminosity functions are shown in
Figure~\ref{fig:lum-funcs} and Table~\ref{tab:lf-non-para}.  To fit to these
luminosity functions, we increase the errors slightly in order to avoid being
overly constrained by the high S/N bins, which could be dominated by systematic
errors and in consideration of large-scale structure uncertainties.  The
results of fitting Schechter functions are shown in Figure~\ref{fig:lum-funcs}
and Table~\ref{tab:lf-schechter}.  Overall, about 42\% of the $r$-band
luminosity density is in red-distribution galaxies.  Not surprisingly, this is
slightly larger than the 38\% found to be in red-type galaxies by
\cite{hogg02red} because their definition of red-type was based on strict cuts
in color, concentration and surface brightness.

%% table values may change if photo version or k-corrections are changed %%

\begin{table} %[f]
\caption{Luminosity functions\rlap{$^a$}}
\label{tab:lf-non-para}
\begin{center}
\begin{tabular}{ccc}
\hline
$M_r - 5 \log h_{70}$ & $\phir\,h_{70}^{-3}$ & $\phib\,h_{70}^{-3}$ \\ 
& (${\rm mag}^{-1}\,{\rm Mpc^{-3}}$) & (${\rm mag}^{-1}\,{\rm Mpc^{-3}}$) \\ 
\hline
$-23.25$ & $(2.06\pm0.35)\times10^{-5}$ & $(0.30\pm2.83)\times10^{-6}$\\
$-22.75$ & $(0.99\pm0.07)\times10^{-4}$ & $(2.45\pm0.44)\times10^{-5}$\\
$-22.25$ & $(3.30\pm0.15)\times10^{-4}$ & $(2.04\pm0.12)\times10^{-4}$\\
$-21.75$ & $(5.81\pm0.24)\times10^{-4}$ & $(6.24\pm0.26)\times10^{-4}$\\
$-21.25$ & $(8.82\pm0.33)\times10^{-4}$ & $(1.18\pm0.04)\times10^{-3}$\\
$-20.75$ & $(1.14\pm0.04)\times10^{-3}$ & $(1.68\pm0.06)\times10^{-3}$\\
$-20.25$ & $(1.30\pm0.05)\times10^{-3}$ & $(2.09\pm0.07)\times10^{-3}$\\
$-19.75$ & $(1.27\pm0.05)\times10^{-3}$ & $(2.31\pm0.08)\times10^{-3}$\\
$-19.25$ & $(1.28\pm0.06)\times10^{-3}$ & $(2.85\pm0.10)\times10^{-3}$\\
$-18.75$ & $(1.18\pm0.06)\times10^{-3}$ & $(3.48\pm0.13)\times10^{-3}$\\
$-18.25$ & $(1.12\pm0.08)\times10^{-3}$ & $(4.74\pm0.20)\times10^{-3}$\\
$-17.75$ & $(1.23\pm0.12)\times10^{-3}$ & $(5.56\pm0.30)\times10^{-3}$\\
$-17.25$ & $(1.17\pm0.16)\times10^{-3}$ & $(6.27\pm0.49)\times10^{-3}$\\
$-16.75$ & $(0.95\pm0.22)\times10^{-3}$ & $(5.43\pm0.70)\times10^{-3}$\\
$-16.25$ & $(1.01\pm0.40)\times10^{-3}$ & $(0.99\pm0.24)\times10^{-2}$\\
$-15.75$ & $(1.77\pm0.92)\times10^{-3}$ & $(1.27\pm0.54)\times10^{-2}$\\
\hline
\end{tabular}
\end{center}
{\small
$^a$~Non-parametric luminosity functions for the red and blue distributions
(see Eqn.~\ref{eqn:distribution}). The errors include formal and systematic
uncertainties.  The latter include a constant 3\% plus a fraction proportional
to 1/\Vmax\ increasing to 40\% for the lowest luminosity bin (to approximately
account for large-scale structure effects). The functions are shown in
Figure~\ref{fig:lum-funcs}.
}
\end{table}

%% table values may change if photo version or k-corrections are changed %%

\begin{table*} %[f]
\caption{Schechter function fits to the luminosity functions\rlap{$^a$}}
\label{tab:lf-schechter}
\begin{center}
\begin{tabular}{ccccccc}
\hline
distribution & $M^{*} - 5 \log h_{70}$ & $\phi^{*}\,h_{70}^{-3}$ 
& $\alpha$ & $\phi^{*}{}'\,h_{70}^{-3}$ & $\alpha'$ 
& $j + 2.5 \log h_{70}$\rlap{$^b$} \\
& & ($10^{-3}\,{\rm Mpc^{-3}}$) & 
& ($10^{-3}\,{\rm Mpc^{-3}}$) & & (${\rm Mpc^{-3}}$) \\
\hline
$\phir$&$-21.49\pm0.03$&$2.25\pm0.08$&$-0.83\pm0.02$
&---&---&$-14.79$ (42\%)\\
$\phib$&$-20.60\pm0.08$&$2.82\pm0.32$&$+0.26\pm0.21$
&$2.35\pm0.37$&$-1.35\pm0.05$&$-15.13$ (58\%)\\
$\phib$&$-21.28\pm0.03$&$2.89\pm0.13$&$-1.18\pm0.02$
&---&---&$-15.08$ \\
\hline
\end{tabular}
\end{center}
{\small
$^a$~A single Schechter function was found to give a good fit to the red
distribution ($\phir$) but not to the blue distribution ($\phib$). In the
latter case, a significantly better fit was obtained by summing two Schechter
functions (with a single value for $M^{*}$). Both the double- and
single-Schechter function parameters are shown for $\phib$.  \vspace{0.1cm}
\newline %-
$^b$~The luminosity density in absolute magnitudes per Mpc$^3$.  The
percentage in brackets is the fraction relative to the total $r$-band
luminosity density.
}
\end{table*}

Note that a single Schechter function was found to give a good fit to the red
distribution but not to the blue distribution. In the latter case, there is a
small but statistically significant slope change around $M_r\approx-20$. To
account for this, we used a double-Schechter function but with the same value
for $M^{*}$ (i.e.\ the sum of two power laws with one exponential cutoff, it
was not necessary to allow two-different $M^{*}$ values to provide a good
fit). The double-Schechter function provided a noticeably better fit to the
faint end of the luminosity function with a steeper faint-end slope
($\alpha'=-1.35$, the second power law dominates here) compared to the single
Schechter fit ($\alpha=-1.18$). Note that this is purely a mechanism for
obtaining a better fit to the luminosity function and it should {\em not} be
interpreted as evidence for two blue populations.

The red distribution has a significantly shallower faint-end slope
($\alpha=-0.83$) than the blue distribution.  Related results have been found
by dividing galaxies into classes from early to late spectral types:
\cite{madgwick02} found faint-end slopes from $-0.5$ to $-1.5$ based on
emission and absorption line strengths in optical spectra, and;
\cite{blanton01} found a steepening of the slope from red to blue galaxies
based on cuts in $g-r$ color (their fig.~14).  However, the equivalent
steepening of the faint-end slope toward late types based on morphological
classification appears less significant \citep{nakamura03}.  This is not
inconsistent with our result since the red and blue distributions at the faint
and bright ends need not have the same mix of morphological types.  In other
words, the processes that result in the red (or the blue) distribution also
produce a range of morphological types that need not be the same at low and
high luminosities.

\subsection{Dividing the Distribution}
\label{sec:divide}

\begin{figure} %[f]
\epsscale{1.1}\plotone{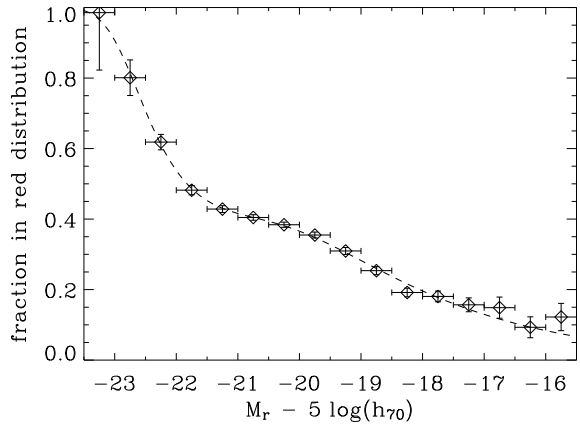}
\caption{Fraction of galaxies that are part of the red distribution{} 
  as a function of absolute magnitude: $\phir(M_r)$ / [$\phir(M_r)+
  \phib(M_r)]$.  The {\em diamonds} with error bars represent the ratios
  determined from the non-parametric luminosity functions
  (Table~\ref{tab:lf-non-para}) while the {\em dashed line} represents the
  ratios determined from the Schechter function fits
  (Table~\ref{tab:lf-schechter}).}
\label{fig:relative-red}
\end{figure}

One of the ways to divide a galaxy sample is by absolute magnitude to compare,
for example, galaxy clustering relations \citep{zehavi02}.
Figure~\ref{fig:relative-red} shows the ratio between the luminosity functions
as a function of absolute magnitude. The ratio of the red distribution to the
total galaxy population gradually increases from low luminosities to
$M_r\approx-22$. For galaxies more luminous than this, the fraction of the
population derived from the red distribution increases more rapidly.  This
agrees with the standard result that the most luminous galaxies are almost
entirely early types \citep[e.g.][]{blanton03broadband}.

\begin{figure} %[f]
\epsscale{1.1}\plotone{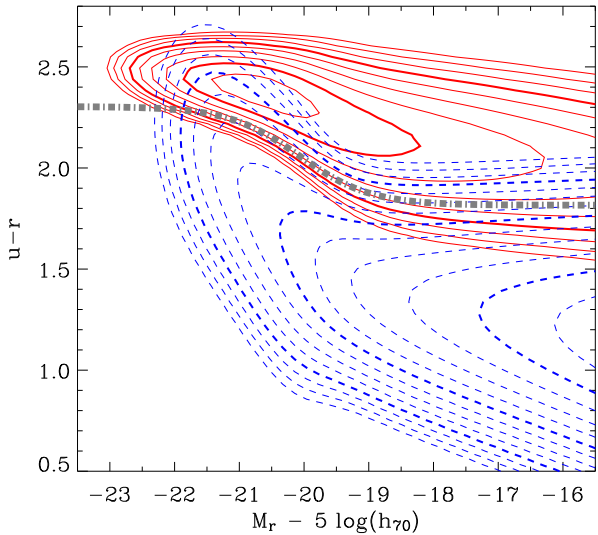}
\caption{Empirical red and blue bivariate distributions{}
  ($\Phi_{\rm r}$ and $\Phi_{\rm b}$) using the parameterization of
  Section~\ref{sec:parameterization} with values from
  Tables~\ref{tab:t-params} and~\ref{tab:lf-schechter}. The {\em solid
    contours} represent the red distribution, while the {\em dashed contours}
  represent the blue distribution.  The contour levels are on a logarithmic
  scale, starting at an arbitrary level and doubling every two contours. The
  {\em thick dash-and-dotted line} represents an optimal divider
  (Sec.~\ref{sec:divide}). Note that the measured dispersion includes
  observational uncertainties, which implies that the luminous red `ridge' is,
  in reality, significantly narrower.}
\label{fig:bimodal-bivariate}
\end{figure}

We can also look at divisions in color and absolute magnitude space.
Figure~\ref{fig:bimodal-bivariate} shows the bivariate distribution separated
into the two types based on the analysis in this paper. In general, it is
possible to define regions of the space that are almost entirely derived from
one distribution except, notably, for a region around $M_r\sim-21.5$ and
$C_{ur}\sim2.5$. Here, dusty and/or bulge-dominated spirals `overlap' in this
space with old stellar-population ellipticals. A related result was obtained
by \cite{hogg03}, studying ``the overdensities of galaxy environments as a
function of luminosity and color'', where luminous red and faint red galaxies
were found, on average, in more overdense regions than $M^{*}$ red galaxies
(see their fig.~2).  Our analysis can explain this result because of the blue
distribution interlopers, formed by a different set of processes, with similar
colors to the $M^{*}$ red distribution galaxies.

Modeling a distribution with two sequences in this way naturally leads to a
physical model with two kinds of galaxies with different processes associated
with them. Therefore, it would be appropriate to study the properties of each
distribution separately. However, this is not precisely possible because of
the dispersions and overlaps associated with each distribution. Instead, we
can make an optimal divider by defining figure of merits based on the
double-Gaussian description. Following \cite{strateva01}, for any cut on
color, we can estimate the `completeness' (${\cal C}$) and `reliability'
(${\cal R}$) of a sample. For example, if we use $C_{ur}>C'_{ur}$ to select
the red distribution then: $\comr$ is the fraction of galaxies from the red
distribution that are selected, and; $\relr$ is the fraction of galaxies
selected that derive from the red distribution (i.e.\ $1-\relr$ is the
contamination from the blue distribution).

%% divider coeffs. may change if photo version or k-corrections are changed %%

There are many ways to define an optimum divider in color as a function of
absolute magnitude based on different weightings of completeness and
reliability. They can be determined using the parameterized description of the
data described in this paper. Here, we define an optimum divider that best
selects red distribution galaxies redder than the color cut and vice versa
simultaneously, with a figure of merit defined by $\comr\relr\comb\relb$. This
optimal divider (parameterized by a tanh function) is given by
\begin{equation}
 C'_{ur}(M_r) = 2.06 - 0.244 \tanh\left[ \frac{M_r + 20.07}{1.09} \right]
\label{eqn:divide}
\end{equation}
and is shown in Figures~\ref{fig:mean-var} and~\ref{fig:bimodal-bivariate}.
The optimal color division varies from about 2.3 at the bright end to 1.8 at
the faint end of the galaxy distribution. For galaxies fainter than $M_r$ of
$-21$, we obtain $\comr>0.8$, $\relr>0.8$, $\comb>0.85$ and $\relb>0.95$ at
all magnitudes, but for more luminous galaxies, both $\comb$ and $\relr$ drop
below 0.8 due to the increased overlap of the blue distribution with the red
distribution ($\relr$ rises again for the most luminous two or three bins
because of the thinning out of the blue distribution).

\subsection{Conversion to Stellar Mass}
\label{sec:stellar-mass}

In terms of relating variations in galaxy properties to models of galaxy
formation and evolution, it is more appropriate to consider stellar mass than
luminosity because stellar mass is more closely related to baryon content.
Stellar mass-to-light ratios (\ml) can vary by up to a factor of about ten for
the $r$-band luminosity. However, \ml\ can be estimated by fitting
population-synthesis models to colors or spectroscopic indices
\citep{BE00,bell03,kauffmann03A}.  In order to convert our results to
stellar-mass relations, we use an approximate color-\ml\ conversion given by
\begin{equation}
 \log ({\cal M}/{L_r}) \:=\: a \, + \, b \, C_{ur}
\label{eqn:M/L}
\end{equation}
where $(a,b)=(-0.55,0.45)$ and; ${\cal M}$ and $L_r$ are the mass and
specific luminosity in solar units.\footnote{We use ${\cal M}$ for mass and
  $M$ for absolute magnitudes.  The conversion is given by $\log ({\cal M} /
  {\cal M}_{\odot}) = (M_{r\odot} - M_r)/2.5 + \log ({\cal M}/{L_r})$ where
  $M_{r\odot} = 4.62$ \citep{blanton01}.} This is a useful approximation
because there is a significant correlation between $u-r$ and \ml.

The coefficients in Equation~\ref{eqn:M/L} were derived from an average of
analyses based on the stellar masses of \cite{bell03} and \cite{kauffmann03A},
for which we obtained $(a,b)\approx(-0.3,0.35)$ and $(-0.8,0.55)$,
respectively, by fitting log (stellar mass) as a function of $(u-r)_{\rm
  model}$ for low redshift galaxies ($z<0.08$).  The assumed stellar IMFs were
similar between the two analyses,\footnote{\citeauthor{bell03}\ used a `diet'
  \cite{Salpeter55} IMF, which gives about 70\% of the \ml\ compared to a
  `standard' Salpeter IMF, and \citeauthor{kauffmann03A}\ used a
  \cite{kroupa01} IMF (eqn.~2 of that paper).  These IMFs were found to be
  consistent with cosmic SFH and luminosity densities \citep{BG03}, i.e.\ with
  average galaxy colors, and with galaxy rotation curves \citep{BdJ01}.} and
therefore the differences arise principally from the methodologies (see
\citeauthor{bell03}\ and \citeauthor{kauffmann03A} for details).  This gives
us some estimate of the systematic uncertainties involved with this type of
modeling.  For the stellar-mass ranges quoted in this section (below), we use
the $a,b$ coefficients given above and include uncertainties from our fitting.
Note that we do not include uncertainties in the stellar IMF (or, e.g.\ 
evolutionary tracks), which could amount to $\sim$30\% uncertainty in the
absolute values of the stellar masses, and the conversion to total mass is
considerably more uncertain due to the dominance of dark matter in most
galaxies.

\begin{figure} %[f]
\epsscale{1.1}\plotone{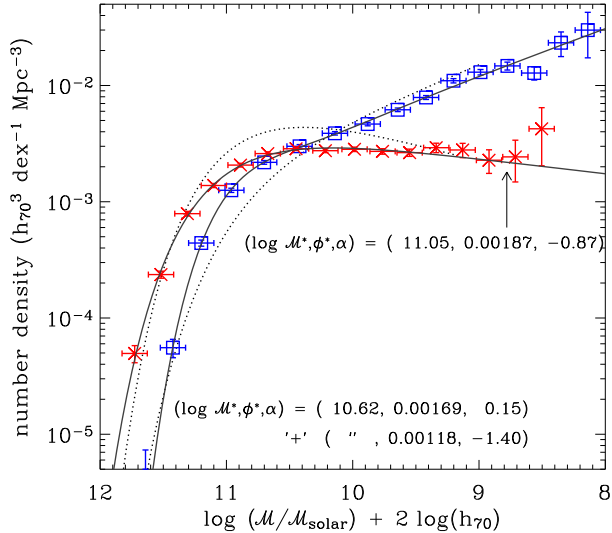}
\caption{Galaxy stellar mass functions,{}
  for each distribution, derived from luminosity functions adjusted to account
  for variations in the stellar mass-to-light ratio as a function of color
  (Eqn.~\ref{eqn:M/L}).  Note that, as well as the $x$-axis being adjusted,
  the number density is adjusted to account for the stretching of the
  magnitude bins and the conversion to base-ten logarithms.  The {\em solid
    lines} represent the Schechter functions fits to our data while {\em
    dotted lines} represent the color-selected early- and late-type GSMF fits
  of \cite{bell03}.  See Figure~\ref{fig:lum-funcs} for other details.}
\label{fig:lum-funcs-adj}
\end{figure}

For simplicity, we apply the \ml\ adjustment (Eqn.~\ref{eqn:M/L}) to the
relations and luminosity functions using the $\muer$ and $\mueb$ values as a
function of absolute magnitude (Table~\ref{tab:t-params}). This is a
reasonable adjustment for the average galaxies in each distribution.
Figure~\ref{fig:lum-funcs-adj} shows the luminosity functions adjusted for
stellar mass-to-light ratios, in effect, galaxy stellar mass functions
(GSMFs). The parameters for the Schechter fits are shown in the plot.  The red
distribution is shifted to higher masses with respect to the blue
distribution. The stellar mass density per magnitude is dominated by the red
distribution for galaxy stellar masses greater than about
(2--5)\,$\times10^{10}$\,\Msun. Overall, about 54\%--60\% of the stellar mass
density is in red-distribution galaxies (depending on the coefficients of the
approximate conversion to stellar mass).

In Figure~\ref{fig:lum-funcs-adj}, we also plot the color-selected early- and
late-type GSMFs of \cite{bell03}.  Notably, the early types have a
significantly higher number density per magnitude relative to the late types
around ${\cal M}^{*}$ whereas the GSMFs from the double-Gaussian fitting have
similar number densities here.  This reflects the fact that our analysis
quantifies an overlap in color space (Fig.~\ref{fig:bimodal-bivariate}), and
thus enhances the `late type' number density compared to a standard color
selection, even if using a slope in $g-r$ versus $M_r$ (as per
\citeauthor{bell03}).

The transitions in galaxy properties occur around
(1.5--2.2)\,$\times10^{10}$\,\Msun\ for the red distribution ($q_1$ for
$\muer$ and $\sigr$) and around (2--3)\,$\times10^{10}$\,\Msun\ for the blue
distribution ($q_1$ for $\mueb$), based on converting the CM relations (see
Table~\ref{tab:t-params}). Despite our simplistic treatment of \ml\ 
conversions (which do not differentiate between dust attenuation and SFH
effects), our transition masses are close to the transition in galaxy
properties noted by \cite{kauffmann03B} that occurred around
$3\times10^{10}$\,\Msun.  Here, we have resolved this transition into three
different effects, a change in dominance from one distribution to the other, a
change in the properties of the red distribution and a change in the blue
distribution.

\section{Conclusions}
\label{sec:conclusions}

We have devised a new method of analysing color-magnitude relations based on
considering double-Gaussian distributions in color
(Figs.~\ref{fig:resolve-dg-a} and~\ref{fig:resolve-dg-b}) rather than strict
cuts on morphological or other properties. From this, we obtain CM and
dispersion-magnitude relations for two dominant red and blue distributions,
which can in general be associated with classical definitions of early- and
late-type galaxies. These relations are evident across seven magnitudes
(Figs.~\ref{fig:sigma-var} and~\ref{fig:mean-var}) but are {\em not} well fit
by a straight line.  Instead, we find that a straight line plus a tanh
function provides good fits (Eqn.~\ref{eqn:sl-tanh},
Table~\ref{tab:t-params}).

For both the red and blue distributions, we can associate the general trend
(the straight line part of the combined function) with a universal
metallicity-luminosity correlation.  The tanh function can be associated with
a transition in other properties of the galaxy population, which could include
star-formation history and dust attenuation (in the case of late types).  Note
we have not proved the above physical explanations but have obtained them from
previous results and analyses in the literature
\citep[e.g.][]{ZKH94,KA97,PdG98,tully98,Garnett02,kauffmann03B}.  Further work
is required, e.g.\ population synthesis fitting to SDSS spectra, to bolster
and quantify the physical explanations for these relations.

After converting to stellar mass, we find that the midpoints of the
transitions parameterized by the tanh functions are around
$2\times10^{10}$\,\Msun\ (Table~\ref{tab:t-params}). In addition, we find that
the number density per magnitude of the red distribution overtakes the blue
distribution at about $3\times10^{10}$\,\Msun\ (Fig.~\ref{fig:lum-funcs-adj}).
These changes in properties of the galaxy population are in good agreement
with the transition found by \cite{kauffmann03B} at $3\times10^{10}$\,\Msun\ 
using spectroscopic measurements.

In order to study the physical properties of each distribution separately, it
is necessary to divide them. To do this, we defined an optimum divider based
on minimizing the overlap between the two Gaussian descriptions
(Eqn.~\ref{eqn:divide}, Fig.~\ref{fig:bimodal-bivariate}).  We note that this
works well for galaxies fainter than $M_r\sim-21$. For galaxies more luminous
than this, morphological indicators that also show a bimodality can work
better at dividing the population into two types. Thus, a weighted
combination of various measurements (from photometry, spectroscopy and
morphology) could provide a better division by type, with the weights varying
with absolute magnitude.

The luminosity functions of the two distributions are significantly different
from each other (Figs.~\ref{fig:lum-funcs} and~\ref{fig:relative-red},
Tables~\ref{tab:lf-non-para} and~\ref{tab:lf-schechter}). The red distribution
luminosity function has a shallower faint-end slope and a more luminous
characteristic magnitude. The difference between the two distributions can be
explained in terms of a merger scenario where the red distribution derives
from more major mergers.  To show this, we first approximately converted the
luminosity functions to galaxy stellar mass functions
(Fig.~\ref{fig:lum-funcs-adj}) and fitted a simple numerical model to the data
(Fig.~\ref{fig:merge-fit}).  Some discussion of mergers and a description of
the model is given in the Appendix.  This is consistent with hierarchical
clustering theories.

\begin{figure} %[f]
\epsscale{1.1}\plotone{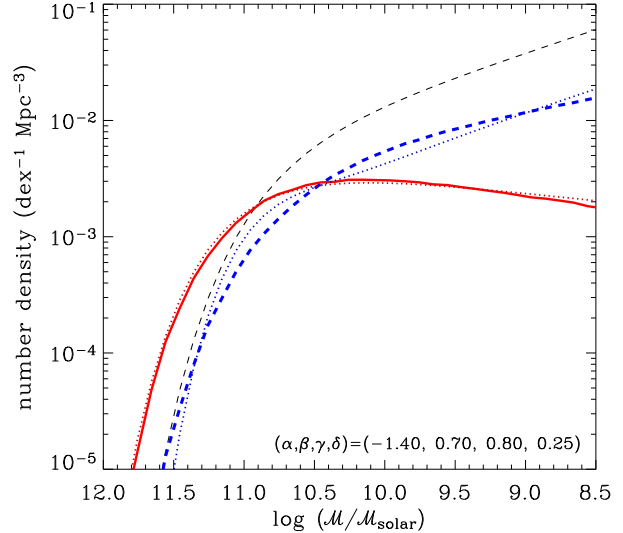}
\caption{Simulated galaxy stellar mass functions,{} 
  for each distribution, from a simple merger model ({\em solid line} and {\em
    thick dashed line} for the red and blue distributions, respectively).  See
  the Appendix for details of the model.  The merger scenario produces a
  shallower faint-end slope and a more massive characteristic mass for the red
  distribution compared to the blue distribution.  The {\em dotted lines}
  represent the Schechter function fits to our data
  (Fig.~\ref{fig:lum-funcs-adj}) while the {\em thin dashed line} represents
  the initial function in the model. The parameters for the model are shown in
  the plot and relate to: initial faint-end slope ($\alpha$); mass exponent
  for weighting in the merger model ($\beta$); probability of a faint galaxy
  merging with a more massive galaxy ($\gamma$), and; fractional increase in
  mass from merging used to determine which galaxies are part of the red
  distribution ($\delta$).}
\label{fig:merge-fit}
\end{figure}

Finally, we note that further work could proceed in a number of directions
including: (i) defining an optimum division between the two distributions by
combining various observed quantities; (ii) analysing the spectra of each
distribution; (iii) studying the distributions of the morphological
properties; (iv) comparing the CM relations between different galaxy
environments, and; (v) simulating galaxy mergers and hierarchical clustering
to test the cause of the bimodality.  Here, we propose that the
double-Gaussian fitting technique represents a model-independent way of
defining a `post-major-merger' sequence, in that the uncertainties due to
blue-distribution interlopers are quantified, without using a semi-arbitrary
cut in morphology or spectral type.

\acknowledgements{We thank the anonymous referee, Eric Bell, Tamas Budavari,
  Tomo Goto, Timothy Heckman, Rachel Somerville and Christina Tremonti for
  helpful suggestions and discussion, and Michael Blanton, Daniel Eisenstein,
  David Hogg, Jeffrey Munn, Adrian Pope, David Schlegel, Max Tegmark and Idit
  Zehavi for maintaining the NYU LSS samples. We acknowledge NASA's
  Astrophysics Data System Bibliographic Services and the IDL Astronomy User's
  Library as valuable resources. IKB and KG acknowledge generous funding from
  the David and Lucille Packard foundation.

  Funding for the creation and distribution of the SDSS Archive has been
  provided by the Alfred P. Sloan Foundation, the Participating Institutions,
  the National Aeronautics and Space Administration, the National Science
  Foundation, the U.S. Department of Energy, the Japanese Monbukagakusho, and
  the Max Planck Society. The SDSS Web site is {\tt http://www.sdss.org/}.

  The SDSS is managed by the Astrophysical Research Consortium (ARC) for the
  Participating Institutions. The Participating Institutions are The
  University of Chicago, Fermilab, the Institute for Advanced Study, the Japan
  Participation Group, The Johns Hopkins University, Los Alamos National
  Laboratory, the Max-Planck-Institute for Astronomy (MPIA), the
  Max-Planck-Institute for Astrophysics (MPA), New Mexico State University,
  University of Pittsburgh, Princeton University, the United States Naval
  Observatory, and the University of Washington.
}

\appendix

\section{A Simple Merger Model}
\label{sec:mergers}

Early-type galaxies tend to have more spherical geometries, more virialized
motions of stars, less dust as well as redder colors. N-body simulations
suggest that the geometries and the motions of stars, similar to those
observed in ellipticals, can be produced by galaxy mergers
\citep{Barnes88,BH92}.  In addition, if the merger causes the gas and dust to
be expelled and/or used up in a burst of star formation \citep{JW85} then the
galaxy's star-formation rate will be lower (at some later time) than
star-forming late-type galaxies. This in turn will mean redder colors for
galaxies produced by mergers as long as any induced star burst does not
dominate the stellar population or the merger occurred at high redshift, i.e.\ 
as long as most of the stars formed at high redshift
\citep*{BCF96,Kauffmann96}.  \cite{KC98} have shown that a hierarchical merger
model can reproduce the CM relation of cluster ellipticals.

Given these lines of argument, it is reasonable to suppose that mergers are
the cause of the bimodality, with the red distribution deriving from major
merger processes and the blue distribution deriving from more quiescent
accretion (with only minor mergers at most).  To test this, we devised an
illustrative non-dynamical merger model to see if the basic shapes of the
GSMFs (Fig.~\ref{fig:lum-funcs-adj}) with respect to each other could be
explained.  The procedure for this model is described below.
\begin{enumerate}
\item A population of galaxies is created with an initial baryonic-mass
  function described by a Schechter function with a faint-end slope $\alpha$
  (for simplicity, we assume that all the baryons will be used to form stars
  and thus can be related to the GSMFs observed today).  The population is
  defined from about $10^{-3}\,{\cal M}^{*}$ to $10\,{\cal M}^{*}$. The
  characteristic mass and number density, ${\cal M}^{*}$ and $\phi^{*}$, are
  adjusted to best match the data after the simulation.
\item These galaxies are numbered from 1 to $N_{\rm gals}$ in order of
  increasing mass.
\item For each galaxy $i$, it is determined whether it will merge with a {\em
    more massive} galaxy based on a probability equal to
\begin{equation}
  p_i = \left( \frac{\sum_{j=i+1}^{N_{\rm gals}} {\cal M}_j^\beta}
    {\sum_{j=1}^{N_{\rm gals}} {\cal M}_j^\beta} \right) \gamma
\end{equation}
where ${\cal M}_j$ is the initial mass of the $j$-th galaxy.  In other words,
the probability is the sum of the more massive galaxies weighted by mass with
an exponent $\beta$, divided by the total mass in the population, multiplied
by $\gamma$.  The probability of the lowest mass galaxy merging with another
galaxy is approximately $\gamma$.
\item For each merged galaxy $i$, the mass is added to another galaxy at
  random but with a weighting proportional to ${\cal M}_j^\beta$ for $j>i$
  (and 0 for $j<i$).
\item For each remaining galaxy, the fractional increase of its mass relative
  to its initial mass is determined. Galaxies with fractional increases
  greater than $\delta$ are determined to be in the red distribution
  \citep*[similar to the $f_{\rm ellip}$ parameter of][]{KWG93}.
\item The model GSMFs for the red and blue distributions are determined and
  ${\cal M}^{*}$ and $\phi^{*}$ are adjusted to best fit the data, over the
  ranges $8.6<\log{\cal M}^{*}<11.8$ for the red and $8.4<\log{\cal
    M}^{*}<11.6$ for the blue distribution.
\end{enumerate}
The additional physical assumptions behind this scenario are: that galaxies
form from quiescent accretion with a distribution in masses defined by a
Schechter function, and; the probability of merging with a more massive galaxy
is related to the number density and masses of all these galaxies.  The model
is simple in the sense that it is non-dynamical, the timing of accretion and
merging is not accounted for, and the parameter $\beta$ hides the complex
physics associated with forces on dark-matter haloes and their baryon
contents. Note also that we do not model the CM relations, only the GSMFs.

Figure~\ref{fig:merge-fit} shows a best-fit example of the simulated GSMFs
from this simple merger model.  It reproduces the shape of the
red-distribution GSMF with high accuracy and the approximate faint-end slope
of the blue distribution (though the shape is slightly different).  Thus, the
different luminosity functions (or GSMFs) can be explained if the red
distribution is derived from galaxies where more than a certain fraction of
their mass has come from mergers rather than `normal' quiescent accretion.  In
other words, the red distribution is a post-major-merger sequence where
`major' is determined by the ratios of the masses of the merging galaxies.
This sequence could also include galaxies derived from the sum of many minor
mergers, which could evolve a galaxy from a spiral to an S0 \citep*{WMH96}.

\end{document}